\documentclass[prmat,amsmath,twocolumn,amssymb,floatfixng,showpacs,superscriptaddress,footinbib]{revtex4-1}
\pdfoutput=1
\usepackage[dvips]{graphics}
\usepackage{hyperref}
\usepackage{bm}
\usepackage{epsfig}
\usepackage{enumerate}
\usepackage{color}
\usepackage{graphicx}
\usepackage{subfigure}
\usepackage{txfonts}
\usepackage[dvipsnames]{xcolor}
\usepackage{sidecap,tikz}
\usepackage{graphicx}
\usepackage{makecell}
\usepackage{hyperref}
\usepackage{multirow}
\definecolor{dred}{rgb}{0.6,0,0}
\hypersetup{
    pdfnewwindow=true,      
    colorlinks=true,       
    linkcolor=magenta,          
    citecolor=blue,         
    filecolor=blue,      
    urlcolor=blue        
}
\newcommand{\bea}{\begin{eqnarray}}
\newcommand{\eea}{\end{eqnarray}}
\newcommand{\beq}{\begin{equation}}  
\newcommand{\eeq}{\end{equation}}

\newcommand\ie{i.e.,\,}

\definecolor{lime}{HTML}{A6CE39}
\DeclareRobustCommand{\orcidicon}{\hspace{-1mm}
	\begin{tikzpicture}
	\draw[lime, fill=lime] (0,0) 
	circle [radius=0.16] 
	node[white] {{\fontfamily{qag}\selectfont \tiny \,ID}};
	\draw[white, fill=white] (-0.0525,0.095) 
	circle [radius=0.007];
	\end{tikzpicture}
	\hspace{-3mm}
}
\foreach \x in {A, ..., Z}{\expandafter\xdef\csname orcid\x\endcsname{\noexpand\href{https://orcid.org/\csname orcidauthor\x\endcsname}
			{\noexpand\orcidicon}}
}

\begin{document}
\title{Thermodefect voltage in graphene nanoribbon junctions}
\thanks{NOTICE: The published version of the article can be found at \url{https://doi.org/10.1088/1361-648X/ac553b}}
\author{Alhun Aydin\orcidA{}}
\affiliation{Department of Chemistry and Chemical Biology, Harvard University, Cambridge, MA 02138, USA}
\author{Altug Sisman\orcidB{}}
\affiliation{Department of Physics and Astronomy, Uppsala University, Box 516, S-751 20 Uppsala, Sweden}
\author{Jonas Fransson\orcidC{}}
\affiliation{Department of Physics and Astronomy, Uppsala University, Box 516, S-751 20 Uppsala, Sweden}
\author{Annica M. Black-Schaffer\orcidD{}}
\affiliation{Department of Physics and Astronomy, Uppsala University, Box 516, S-751 20 Uppsala, Sweden}
\author{Paramita Dutta\orcidE{}}
\affiliation{Theoretical Physics Division, Physical Research Laboratory, Ahmedabad-380009, India}
\affiliation{Department of Physics, Birla Institute of Technology and Science - Pilani, Rajasthan-333031, India}
\affiliation{Department of Physics and Astronomy, Uppsala University, Box 516, S-751 20 Uppsala, Sweden}

\begin{abstract}
Thermoelectric junctions are often made of components of different materials characterized by distinct transport 
properties. Single material junctions, with the same type of charge carriers, have also been considered to  
investigate various classical and quantum effects on the thermoelectric properties of nanostructured materials. 
We here introduce the concept of defect-induced thermoelectric voltage, namely, {\it thermodefect voltage}, in graphene 
nanoribbon (GNR) junctions under a temperature gradient. Our thermodefect junction is formed by two GNRs 
with identical properties except the existence of defects in one of the nanoribbons. At room temperature the thermodefect 
voltage is highly sensitive to the types of defects, their locations, as well as the width and edge configurations of 
the GNRs. We computationally demonstrate that the thermodefect voltage can be as high as $1.7$\,mV/K for $555$-$777$ 
defects in semiconducting armchair GNRs. We further investigate the Seebeck coefficient, electrical conductance, and electronic thermal conductance, and also the power factor of the individual junction components to explain the thermodefect effect. Taken together, our study presents a new pathway to enhance the thermoelectric properties 
of nanomaterials.
\end{abstract}

\maketitle

\section{Introduction}\label{I}

Thermoelectricity is the phenomenon of induced electrical potential difference due to a temperature gradient and 
vice versa. Investigation of various junctions based on different materials has been the conventional route for thermoelectric research\,\cite{te3}. Following the footsteps of the pioneering works by Hicks and Dresselhaus \cite{te1,te2}, subsequent efforts have established that quantum confinement can also be used to enhance the 
thermoelectric properties\,\cite{tebook2013,geo1,geo2,geo3,geo4,geo5,tegnr2010ftc,tebook2014,tenw,optimiz,nunez,topterm}. The 
contributions of nanoscale effects can be isolated and analyzed quite easily when all the components of the 
junctions are made of only a single material, having the same dominant charge carriers but different size or 
shape. Such junctions are gathered under the umbrella name of single-material unipolar thermoelectric junctions\,\cite{tsegrap,tshe}. 

The concept of single-material unipolar junctions was first introduced by Sisman and Müller in 
2004 to isolate and analyze the thermoelectric voltage generated only due to the differences of quantum size effects between the 
junction components\,\cite{tse1}. Since then various single-material unipolar thermoelectric junctions formed by 
the combination of nano-macro\,\cite{tse1,tsef2010a,tsef2012a,tsef2012b,tsef2013a,tsef2013b} or nano-micro 
components have been studied\,\cite{tsef2008a,tsef2008b,tsef2011a,tse5,tsef2018a}, considering 
both classical and quantum size effects\,\cite{tse2,tse4,tsef2014a,tsef2018b,tsegrap,tshe,dsey}. In particular, thermoelectric properties of quantum wells and nanowires using graphene-based materials have become a hot topic in material science in recent years\,\cite{PhysRevB.81.113401,PhysRevB.81.235406,Sevincli2013,tebook2014,chempot,topterm,optimiz,Nozaki2014,Sevincli2014,doi:10.1021/acs.nanolett.8b03406,tedef4,tsegrap,therdefposit,doi:10.1021/acsaem.9b02187}. In these, as well as other nano-scale systems, thermoelectric devices are made by connecting many single junctions to each other such that small nano-scale effects of many single junctions are added up to generate a usable macroscopic voltage. As a direct consequence, improving the thermoelectric properties of single junctions is essential to develop superior thermoelectric devices.

\begin{figure}[t]
\centering
\includegraphics[width=0.34\textwidth]{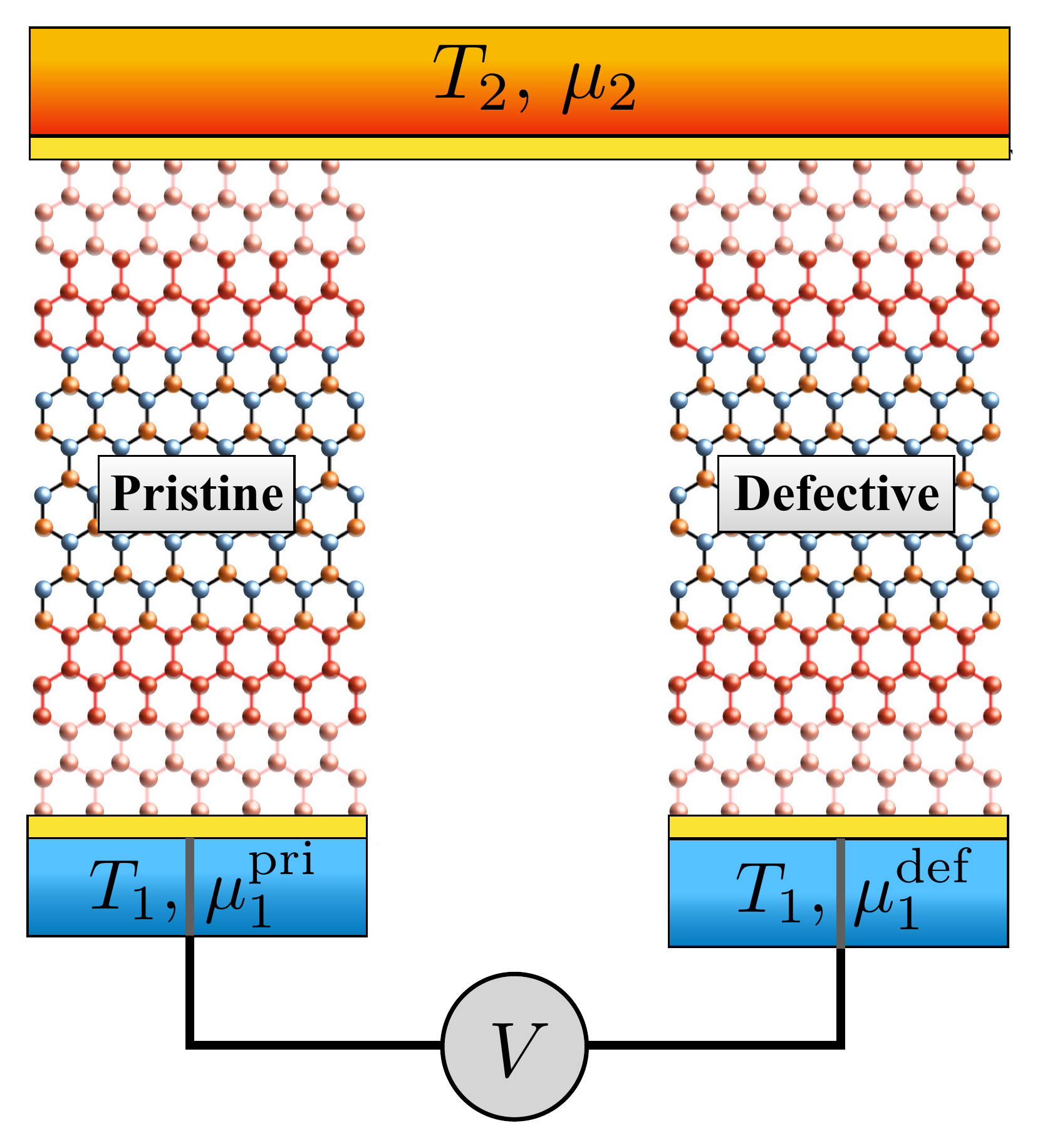}
\caption{Thermodefect junction of pristine and defective GNRs maintained at a temperature difference $T_2-T_1$ 
where $T_2>T_1$ (denoted by orange (hot) and blue (cold) colors). The two GNRs are thermally and electrically 
unified at the top end but electrically separated at the bottom. Yellow bars denote electrically and thermally 
conductive contact materials.}
\label{fig:model}
\end{figure}

In this work, we introduce a new concept: {\it defect-based} single-material unipolar thermoelectric junctions, 
here consisting of one pristine and one defective nanoribbon maintained at the same temperature gradient, as shown 
in Fig.\,\ref{fig:model}. Under an applied temperature gradient, we achieve an (electro-)chemical potential difference 
at the separated end of the junction, which is utilized as thermoelectric voltage. The voltage is generated {\it only} 
due to the existence of defects in one of the junction components, as all other properties of the junction materials 
are exactly the same, which is completely different from the conventional thermoelectric junctions. We call this junction 
a `{\it thermodefect junction}' and the voltage a `{\it thermodefect voltage}'. The names are justified by the origin of the 
voltage i.e.~the thermal gradient and defects in the system. 

To establish the concept of the thermodefect voltage, we choose graphene nanoribbons (GNRs) where both GNRs 
have exactly the same size (both length and width) and edge configurations, the only difference being the existence of crystallographic defects in one of the GNRs. The pristine GNR then acts as the reference component with respect to the defective GNR, in order to isolate and analyze the effect of different defects on the thermoelectric properties. Note that these thermodefect junctions can also be operative when both the GNRs are made defective in different ways as it is the difference in the defective conditions of the ribbons that is essential. However, the target here is to examine the 
effects of various defects, particularly on the thermodefect voltage, and we, therefore, choose a pristine GNR as the 
reference component. Needless to say, in the absence of any defect or temperature difference, the voltage is always zero.

We here use the term {\it defect} to refer broadly to vacancies, such as single vacancy, divacancy, antidot vacancy, line vacancy, as well as impurities and dislocations. With the advancement of nanotechnology, it is nowadays possible 
to engineer a variety of defects in GNRs using tailored electron beam exposure technique\,\cite{expdef0,expdef1,expdef2,expdef3,PhysRevB.78.233407}. In particular, by tuning the electron beam current density and the exposure time, the defect regions can be confined within a very small region of graphene\,\cite{expdef2}. Moreover, these defect regions can be probed without creating further defects\,\cite{expdef0,kelly1998,ouyang2001}. This successful confinement of the defect regions enables the possibility of creating very specific defect structures in graphene. These already realized possibilities of engineering the disorderness \,\cite{doi:10.1021/acs.nanolett.8b03406,tedef1,Haskins2011,PhysRevB.95.155405,tedef2,tedef3,tedef4,doi:10.1021/acsami.0c02155,doi:10.1021/acsaem.9b02187} clearly improves the potential of the present work. 

For a given temperature difference we demonstrate that the thermodefect voltage is sensitive to defect configurations, 
degree of disorder, positions of the defects (in particular for line vacancies), widths of the ribbons, and range of 
chemical potential. Note that the chemical potentials can be tuned by applying a gate voltage to the leads\,\cite{PhysRevLett.105.176602,chempot2}. Specifically, in order to make our results comprehensive, we present the thermodefect voltage for many different defects in GNRs using three different widths of the GNRs that are characterized 
by either metallic or semiconducting behavior based on their edge configurations: armchair GNRs (AGNRs) and zigzag GNRs (ZGNRs). In semiconducting AGNR junctions with Stone-Wales or 555-777 defects, we find thermodefect voltages as large as 1.45 mV/K and 1.7 mV/K, respectively, at room temperature. This can be compared to the conventional semiconducting\,\cite{SHARP2016} and ferromagnet-superconductor hybrid thermoelectric junctions\,\cite{PhysRevLett.116.097001,duttafs1,duttafs2} that usually produce voltages around sub-mV/K or $\mu$V/K, 
showing the impressive prospects of the present work for thermoelectric applications. 

In order to explain and understand the behavior of the thermodefect voltage, we also investigate the thermoelectric
properties of single GNRs, both in absence and presence of various defects. In addition, these analyses allow us 
to identify which defects result in higher performance in terms of thermoelectricity. Specifically, in order to explore 
the thermoelectric properties, we compute the Seebeck coefficient of each GNR, which is the open circuit voltage 
developed across the junction as a result of the inhomogeneous charge distribution driven by the temperature 
gradient\,\cite{blundell2009concepts}. We also investigate the electrical conductance and electronic thermal 
conductance, which measure the ability of charge and heat transfer, respectively. Having understood the 
thermoelectric transport properties, we examine the performance of the GNRs in terms of the power factor as well. 

The rest of the article is organized as follows. In Sec.\,\ref{II} we describe our model, defect types, and the theoretical formalism. Our numerical results for various thermodefect junctions are presented and discussed in Sec.\,\ref{III}. Finally, we summarize and conclude our findings in Sec.\,\ref{IV}.

\section{Model and Formalism}\label{II}
We consider a junction consisting of a pristine and a defective GNR and apply a temperature difference $T_2-T_1$,
where $T_2>T_1$, as shown in Fig.\,\ref{fig:model}. Here, the two GNRs are in electrical contact with one another via a conductor (yellow structure) which is in contact with the thermal reservoir maintained at temperature $T_2$, while they are electrically disconnected from each other at the other end, which is thermally in contact with the reservoir at $T_1$. We denote the chemical potentials at the separated ends by $\mu_1^{\text{pri}}$ and $\mu_1^{\text{def}}$, corresponding to the pristine GNR and defective GNR, respectively, while the common end is held 
at $\mu_2$. 

Given the set-up with an applied temperature difference, a steady-state thermodefect voltage appears due to 
the presence of the defects in the system. The reason behind the appearance of thermodefect voltage can be 
described as follows. Initially, the junction is in a zero external bias condition. When a temperature gradient is 
applied, charge carriers thermally diffuse from the hot to the cold end within a very short time, and then a chemical potential 
gradient is built up as a result of the inhomogeneity of charge carriers induced by thermal diffusion. Thus, the temperature 
difference destroys the charge homogeneity and induces an electrical field, balancing the thermal driving force 
and preserving the zero net current condition in the steady-state. As a consequence, the chemical potentials at 
the separated ends become different from each other due to the existence of defects in one of the junction 
components, while the connected ends of the junction maintain the same chemical potential. This established 
chemical potential difference is then utilized as the thermodefect voltage.

We describe each GNR of width $W$ by the tight-binding Hamiltonian in the Wannier basis,
\begin{equation}
\label{Eq1}
{\cal H}=-t\sum_{\left\langle i,j\right\rangle}c_i^\dagger c_j+\text{H.c.},
\end{equation}
where the sum runs over all the nearest-neighbor sites $\left\langle i,j\right\rangle$ and $t$ is the nearest-neighbor hopping integral. The operator $c_i^\dagger$ ($c_i$) creates (annihilates) an electron at the $i$-th site. We use the 
KWANT quantum transport software package\,\cite{kwant} to construct the tight-binding Hamiltonian and calculate 
the transmission function for each GNR between the two leads. We set $t=2.7$ eV as it successfully describes the 
electronic properties of GNRs\,\cite{gnr2007a,PhysRevB.66.035412,PhysRevB.81.245402,Neto_RMP}. 

\begin{figure*}[t]
\centering
\includegraphics[scale=0.45]{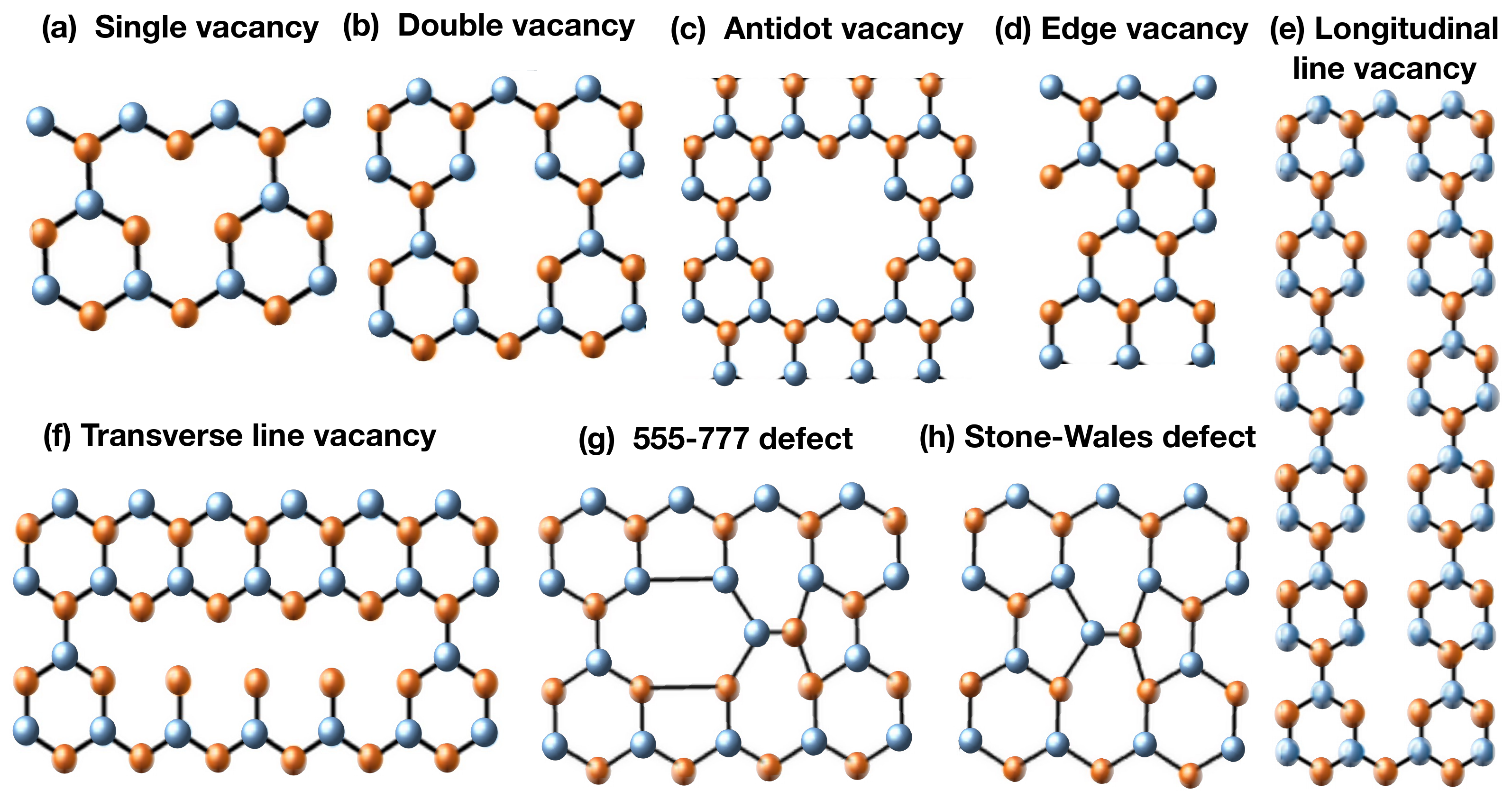}
\caption{Examples of the various defects considered for the defective GNR component of the thermodefect junction. 
The temperature gradient is applied along the vertical direction in each case.}
\label{fig:defects}
\end{figure*}

It is already well-established in the literature that AGNRs of width $W=3n$ and $3n+1$ display semiconducting 
behavior, while they are metallic for widths $W=3n+2$, $n$ being the number of dimers\,\cite{hopsame,Neto_RMP}. 
On the other hand, ZGNRs are always semiconducting with non-dispersive edge states\,\cite{Neto_RMP}. Throughout 
the rest of the work, we use the notations $W$-AGNR and $W$-ZGNR, where $W$ is an integer defining 
the width of each AGNR and ZGNR junction component. We further model the defective GNRs by introducing various types of vacancies, impurities, or dislocations which create disorder in the otherwise regular lattice pattern of the GNR. 
For the calculations here, we consider single, double, edge, antidot and transverse and longitudinal line vacancies, as 
well as Stone-Wales and 555-777 dislocations, all illustrated in Fig.\,\ref{fig:defects}. These represent a wide variety of defects in GNRs that are all stable against the thermal vibrations even at the room temperatures\,\cite{tempdef1,tempdef2,tempdef3}.

To calculate the total density of states (DOS) in the GNRs and how defects influence it, we use the relation $\rho(\varepsilon)=-1/(N \pi) \text{Im}[\text{Tr}[\mathcal{G(\varepsilon)}]]$, where $\mathcal{G(\varepsilon)}=[(\varepsilon+i\eta)\mathcal{I}-\mathcal{H}]^{-1}$ is the Green's function with $\eta$ is an infinitesimal quantity, $\mathcal{I}$ is the Identity operator, and $N$ in the total number of sites in the GNR. 

In the linear regime the net current through a GNR attached to two reservoirs characterized by 
a temperature and chemical potential difference is given by\,\cite{tebook2014}
\begin{align}
I=&-GL\left(\frac{1}{e}\frac{\partial\mu}{\partial x}\right)-GSL\left(\frac{\partial T}{\partial x}\right)
\nonumber\\=&
	-G\left(\frac{\Delta\mu}{e}\right)-GS\Delta T,
\label{Inet}
\end{align}
where $e$ is the electron charge, $L$ is the length along the transport direction, $G$ is the electrical conductance, and 
$S$ is the Seebeck coefficient ($S=-(1/e)(\partial\mu/\partial T)_{I=0}$), while $\partial\mu/\partial x$ and $\partial T/\partial x$ denote chemical potential and temperature gradients respectively. 
The net current should be equal to zero in both components of the junction due to the open circuit condition and this imposes the chemical potentials to be different at the two ends of each GNR. On top of that, the presence of the 
defects in one of the GNRs causes a difference between the chemical potentials at the separated ends of the two 
GNRs. Under zero net current condition, the thermodefect voltage can then be defined by the difference between 
the chemical potentials at the separated ends as
\begin{equation}
V_{\text{TD}}(T_1,T_2,\mu_2)=\frac{1}{e}\left(\mu_1^{\text{pri}}\big|_{I=0}
                                                  -\mu_1^{\text{def}}\big|_{I=0}\right)
\label{Eq3}
\end{equation}
for a given chemical potential $\mu_2$ at the common end. We note that $\mu^{\text{pri}}_1$ and $\mu^{\text{def}}_1$ are to be found in the open circuit condition ($I=0$) using Eq.\,\eqref{Inet}.

Alternatively, Eq.\,\eqref{Eq3} can also be written in the following form
\begin{align}
V_{\text{TD}}=&\frac{1}{e}\int_{T_2}^{T_1}\left[\left(\frac{\partial \mu_1^{\text{pri}}}{\partial T}\right)_{I=0}-\left(\frac{\partial \mu_1^{\text{def}}}{\partial T}\right)_{I=0}\right]dT
\nonumber\\=&
	\int_{T_2}^{T_1}\left(S_{\text{def}}-S_{\text{pri}}\right)dT
	,
\label{Eq4}
\end{align}
which explicitly show the relation between the thermodefect voltage and Seebeck coefficients.
Note that Eq.\,(\ref{Eq4}) reduces to $(S_{\text{pri}}-S_{\text{def}})(T_2-T_1)$ and the Seebeck coefficient also becomes $S=-(1/e)[\mu(T_2)-\mu(T_1)]/(T_2-T_1)$. Thus, Eq.\,(\ref{Eq4}) boils down to Eq.\,(\ref{Eq3}), since $\mu_{\text{pri}}(T_2) = \mu_{\text{def}}(T_2)$ by construction. As a consequence, the thermodefect voltage is directly proportional to the difference in the Seebeck coefficients shown by the defective and pristine GNRs in the 
linear response and ballistic regime. This is already a commonly used approach to calculate thermoelectric properties of GNRs\, \cite{gnr2007a,tegnr2012b,geo3,geo4,tsegrap}. Furthermore, the thermodefect voltage is also linearly proportional to the temperature difference. However, in this study we fix it to $\Delta T=10$ K in order to focus only on the effects of defects.

To examine the Seebeck coefficients, as well as other thermoelectric transport coefficients, we employ the Landauer formalism within the linear response regime. This formalism is described in terms of the transport integrals which read\,\cite{dattabook1995,tebook2014}
\begin{align}
I_\alpha=& \int\left[\beta\left(\varepsilon-\mu\right)\right]^\alpha \beta f(\varepsilon,T)[1-f(\varepsilon,T)]
                  {\cal T}(\varepsilon) d\varepsilon
\end{align}
with $\alpha$ being the energy moment index, $\varepsilon$ the energy, $\cal{T}(\varepsilon)$ the transmission function, and $f(\varepsilon,\mu,T)=1/\{\exp[\beta(\varepsilon-\mu)]+1\}$ the Fermi-Dirac distribution function with 
$\beta=1/(k_B T)$ being the inverse temperature scaled by the Boltzmann constant $k_B$. The thermoelectric transport coefficients: electric conductance $G$, Seebeck coefficient $S$, and electronic thermal conductance $\kappa_e$ are expressed in terms of the transport 
integrals as
\begin{subequations}
\begin{align}
G=& \frac{2e^2}{h}I_0	,\\
S=&	-\frac{k_B}{e}\frac{I_1}{I_0}, \\
\kappa_e=& \frac{2k_B^2}{h}T\left(I_2-\frac{I_1^2}{I_0}\right)
\end{align}
\end{subequations}
where $h$ is the Planck constant. The factor of $2$ appears due to spin degeneracy. The thermoelectric power factor 
$P$ can further be obtained from the relation $P=GS^2$. Note that the ballistic approach allows us to express the non-equilibrium distribution function in terms of equilibrium distribution functions of the reservoirs\,\cite{dattabook1995,qtla}. Therefore, regarding the transport integral, only the transmission function is affected by the existence of defects.

To focus on the influence of the defect difference between the GNR junction components, we assume the electrical contacts to be perfectly transmitting. Small deviations from the fully transmitting condition could make quantitative changes but should not affect the main conclusions of this work. It should also be noted that fixing the chemical potential at the hot or cold end of the junction does not change the results, because the applied temperature difference ($\Delta T=10$~K) is small compared to the set temperature, which is room temperature, $T=300$~K.

As far as interactions are considered, thermoelectric properties are expressed as summations of diffusive (or ballistic), given by Eqs.(6a)-(6c), and interaction terms. Note that the expression of Seebeck coefficient is the same for diffusive and ballistic approaches since it is defined under zero current condition. The contribution of interactions on thermoelectric coefficients is mainly represented by the phonon-drag effect\,\cite{PhysRevB.54.5438,introtebook}. This effect causes an increment of the effective mass of electrons (or holes) as a result of electrons-phonon interactions. Increasing the effective mass decreases the mobility and the conductivity while it increases the Seebeck coefficient due to its inverse proportionality with conductivity. Therefore, effects of phonon interactions on thermodefect voltage needs to be discussed before we continue to use Eqs. (6a)-(6c), because the thermodefect voltage is directly proportional to the Seebeck coefficient. The phonon-drag effect depends on both electron-phonon (e-p), phonon-phonon (p-p) and phonon-boundary (p-b) interactions in general. For graphene, Ref.\,\cite{Bao2010} provides a detailed computational study for the temperature dependency of the phonon-drag contribution on the Seebeck coefficient, ${S_g}$, and its comparison with the diffusive (ballistic here) contribution, ${S}$, Eq.(6b). The model of the study has been verified also by experimental data in the literature. The expression of ${S_g}$ is inversely proportional to the summation of p-p an p-b scattering frequencies (as a result of the Matthiessen's rule\,\cite{ziman}) and also contains a multiplicative term from e-p interaction in the electron relaxation process. The study concludes the following: At low temperatures, the p-p scattering rate decreases as $T^3$ and becomes negligible, while the p-b scattering rate remains almost constant. Therefore, the temperature dependence of ${S_g}$ is mainly due to the e-p process, which causes a temperature dependency in the form of $\exp(-\hbar\Omega_{q,\lambda}/k_BT)/T^2$ for low temperatures where $\Omega_{q,\lambda}$ is the frequency of a phonon of wavevector $q$ and branch $\lambda$. This term has a sharp increase with increasing temperature in the low-temperature regime, around 0.3-2 K. On the other hand, the $T^3$ dependency of the p-p scattering becomes dominant at relatively high temperatures (especially when $T>10$ K) and this process contributes to ${S_g}$ by $T^{-3}$. Also the e-p process gives rise to $T^{-2}$ dependency at high temperatures. Consequently, ${S_g}$ decreases by $T^{-5}$ with increasing temperature in the high temperature regime.  In the same study \,\cite{Bao2010}, it has been shown that the phonon-drag effect becomes negligible and the diffusive term (ballistic here) is essentially the only term contributing to the Seebeck coefficient for temperatures higher than $10$ K. Practically, contributions of interactions become negligible especially when $T>20$ K. There exist also other studies discussing the negligibility of electron-phonon coupling in graphene at room temperature\,\cite{gnr2007a,tegnr2012b}. Since we consider room temperature, we can thus safely neglect contributions of phonons to the Seebeck coefficient as we assume that the existence of defects does not change the main mechanisms suppressing the phonon-drag effect on the Seebeck coefficient. In other words, due to the weak electron-phonon coupling in GNRs, we can ignore phonon effects at room temperature\,\cite{gnr2007a,tegnr2009a,tegnr2012b}.

In the context of phonons and also nuclear vibrations, we additionally mention that effects of vibrational impurities on the electronic structure of graphene have been studied previously\,\cite{PhysRevB.87.245404}. In particular, local vibrations may be pertinent to the vacancies and dislocations that we introduce in the GNRs. However, as was demonstrated in\,\cite{PhysRevB.87.245404}, the Friedel oscillations excited by the local vibrations are fairly short ranged and do not in a significant way distort the overall electronic structure in comparison to the Friedel oscillations already introduced through the presence of the impurities themselves. Hence, our restriction to study ballistic transport should be considered to be firmly founded in the established knowledge of graphene.

\section{Results and Discussions}\label{III}
With the above outline of the theoretical framework, we now present our results of thermodefect voltage and other transport coefficients in two subsections corresponding to two defect classes: vacancies and dislocations. 

\subsection{Vacancies}
\begin{figure}[b]
\centering
\includegraphics[width=0.4\textwidth]{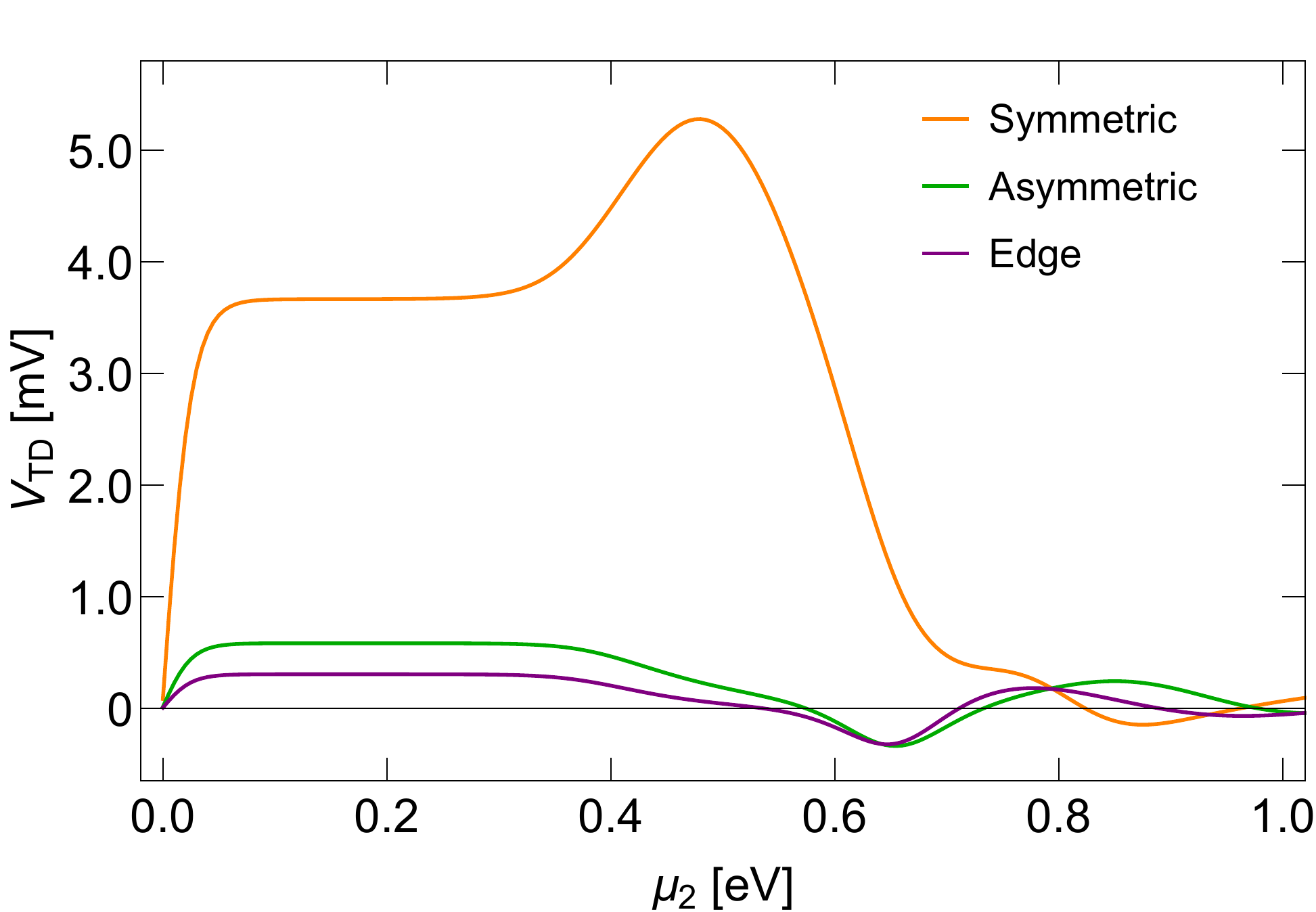}
\caption{Thermodefect voltage for $\Delta T=10$\,K as a function of the chemical potential at the common end 
$\mu_2$ for a thermodefect junction consisting of two $13$-AGNRs. Different colors correspond to the symmetrical (orange), asymmetrical (green) and edge (purple) maximum length longitudinal vacancy in the defective AGNR.}
\label{fig:results_td_v}
\end{figure}

We begin our discussions on thermodefect voltage by considering atomic site vacancies put in a line along the 
longitudinal direction with respect to the transport (see Fig.\,\ref{fig:defects}(e)) in the defective GNR of the 
thermodefect junction of Fig.\,\ref{fig:model}. 
The effect of such a longitudinal line vacancy on the thermodefect voltage strongly depends on the position of the line 
vacancy relative to the edge, which is in agreement with other, non-thermal, transport results in the literature\,\cite{PhysRevB.76.193409,therdef2,therdefposit,PhysRevB.90.115413}. We first in Fig.\,\ref{fig:results_td_v} focus on the behavior of the thermodefect voltage as a function of $\mu_2$ in the presence of longitudinal vacancies at different positions, using AGNRs in the junction. Depending on the position of the line vacancy we call it symmetrical, 
asymmetrical, or edge longitudinal vacancy when the vacancy line is situated in the middle \ie $(W+1)/2$-th line, 
in the $(W+3)/2$-th line, or at the edge of the AGNR, respectively. Note that, exactly symmetric situation 
for the longitudinal vacancy is only possible when $W$ is an odd integer and an example of that is shown in Fig.\,\ref{fig:defects}(e). We can also have line vacancies of different lengths. Here, we only show results for the 
maximum possible length (less than the length of GNR by two atomic sites) in order to achieve the highest 
thermodefect voltages, but the phenomenon of thermodefect voltage is qualitatively similar, except with a reduced magnitude, 
for line vacancies of shorter lengths. 

In Fig.\,\ref{fig:results_td_v} we see that there is a finite thermodefect voltage for a wide range of $\mu_2$ 
including a plateau region up to $\mu_2\sim 0.4$\,eV. This occurs for all three types of the longitudinal line 
vacancies. However, for the symmetrical line vacancy, the induced thermodefect voltage in the plateau 
region is much higher: $3.7$\,mV for $10$\,K, alternatively $0.37$\,mV/K compared to the other two line vacancy positions, and with a maxima as high as $\sim0.5$\,mV/K. For even larger chemical potentials, $\mu_2>0.7$\,eV, the thermodefect voltage drops substantially for all the three types of longitudinal vacancies and show damped 
oscillations around zero \ie both positive and negative voltages are possible for high chemical potentials. These oscillations are associated with the degeneracy effects originated from the Fermi gas properties\,\cite{tebook2014,aydin4,aydin5}. We note here that very high values of the chemical potential, such as $>1$\,eV, are difficult to utilize in practice\,\cite{chempot} and values larger than $2.5$ eV even falls outside the validity of the tight-binding model\,\cite{Neto_RMP}.

To explain the phenomenon of the thermodefect voltage in the presence of longitudinal vacancies and why the 
horizontal position of the vacancy line matters, we draw an analogy with the classical circuit of parallel batteries. A symmetrically divided 
GNR similar to Fig.\,\ref{fig:defects}(e), can be seen as two identical thinner ribbons connected at the two opposite ends, setting up a parallel and short-circuited configuration of a thermosize junction\,\cite{tsegrap}. In this symmetric configuration, the net induced voltage throughout the defective GNR is equal to the voltage of each individual thinner ribbon, since the short-circuit current is zero as a result of the same thermosize voltage induced in each identical ribbon. In the asymmetric case, however, the widths of the two thin GNRs are different, and the defective ribbon starts operating as an individual thermosize cell, which induces a non-zero voltage and short-circuit current in the defective GNR. Since the thinner ribbon has higher resistance, the voltage drop across the thinner GNR becomes higher than that in the thicker ribbon. The total voltage of this parallel and non-identical `battery' circuit finally approaches to the voltage of the thicker ribbon, which is smaller in magnitude (see the Supplementary Material\,\cite{SM} for more details on this analogy). Thus, the largest thermodefect voltage is always induced for a line vacancy causing a symmetric division of the ribbon. 

\begin{figure}[t]
\centering
\includegraphics[width=0.242\textwidth]{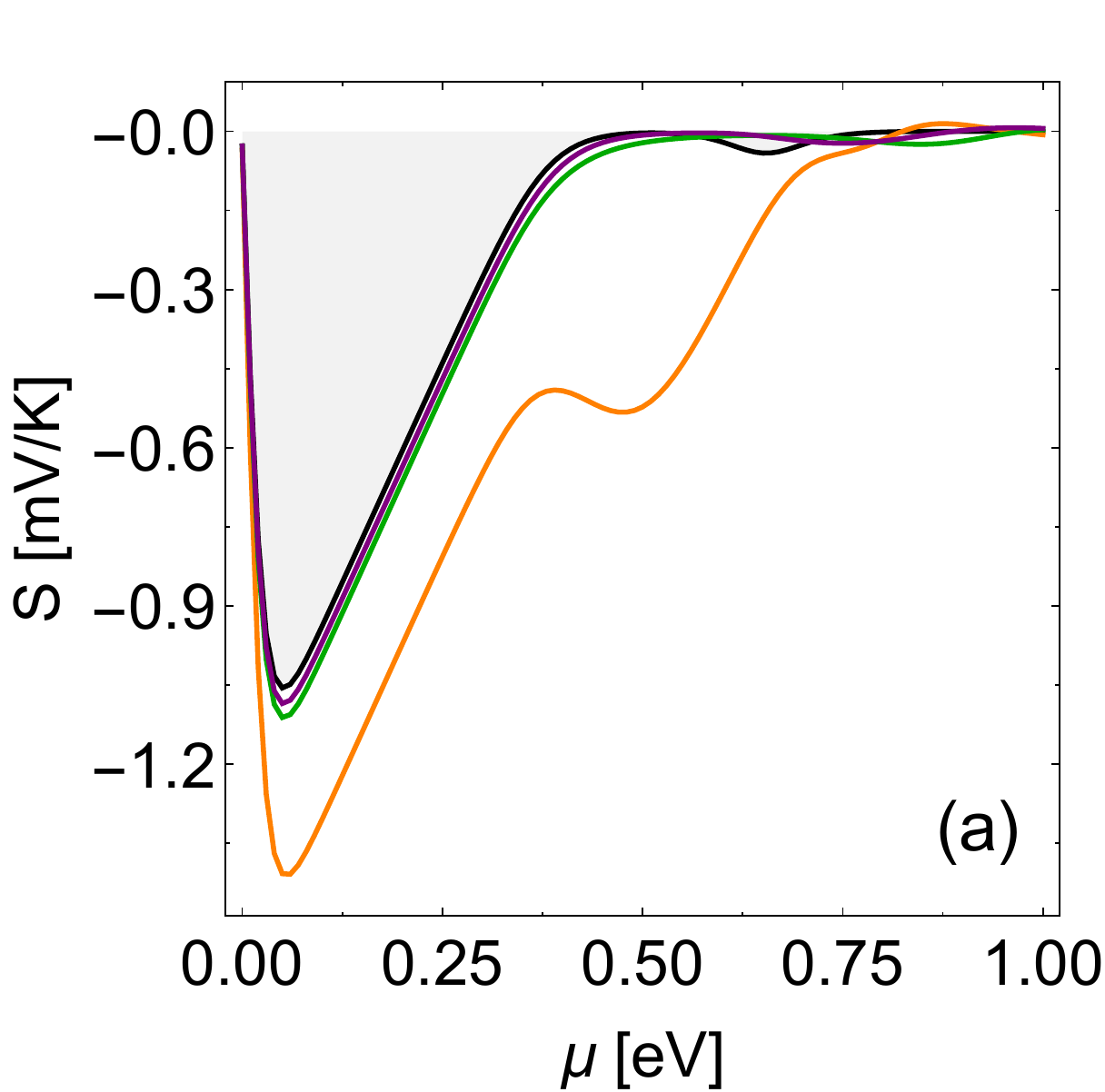}
\includegraphics[width=0.221\textwidth]{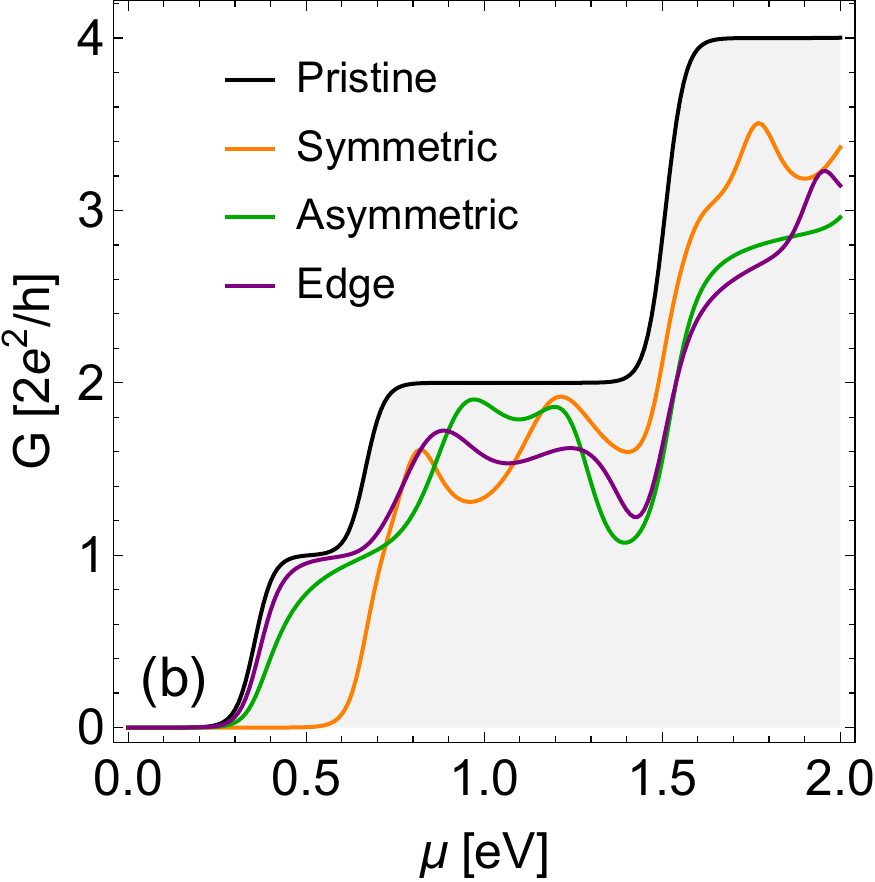}\,\,\,\,\,
\includegraphics[width=0.222\textwidth]{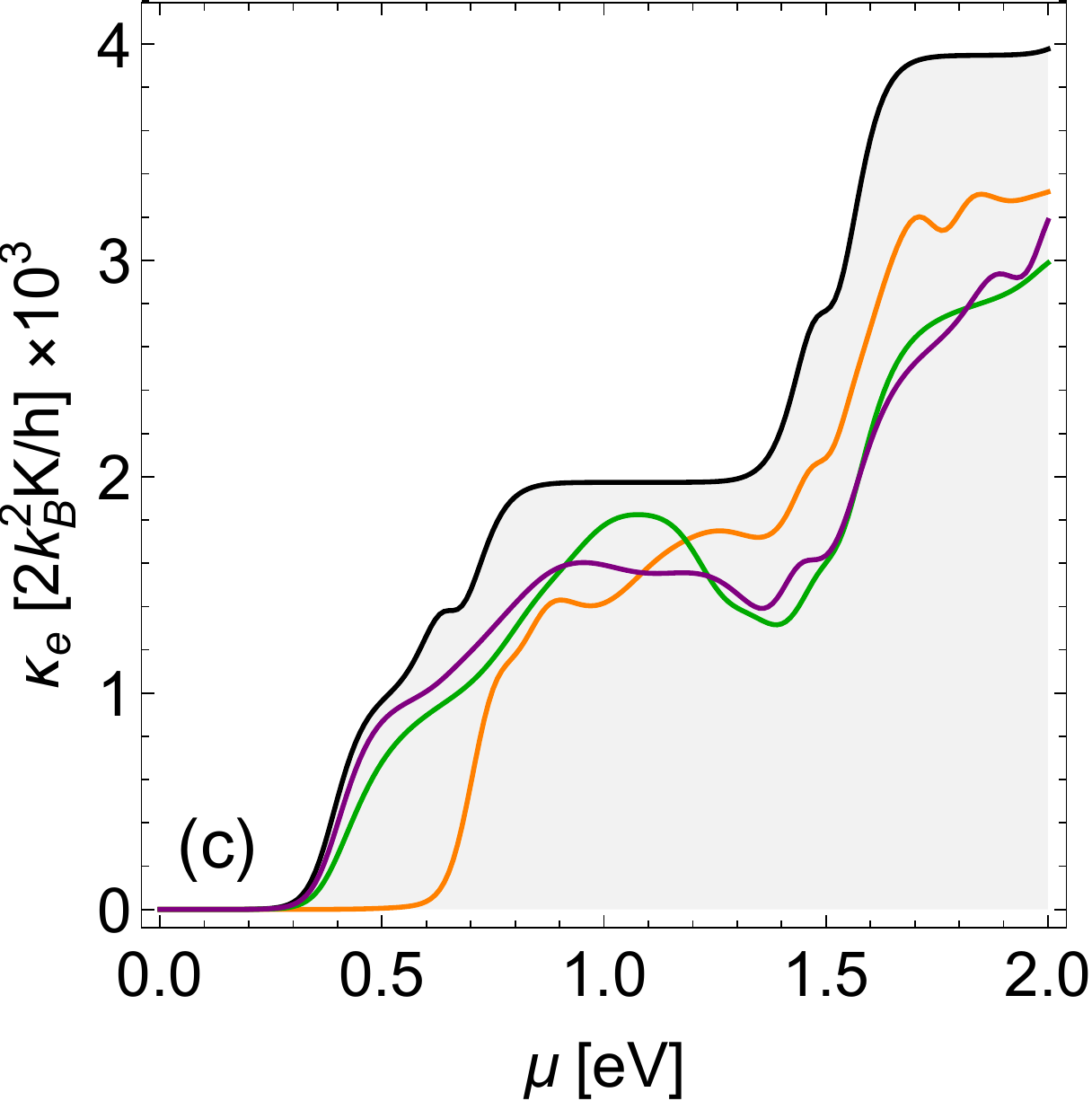}
\includegraphics[width=0.234\textwidth]{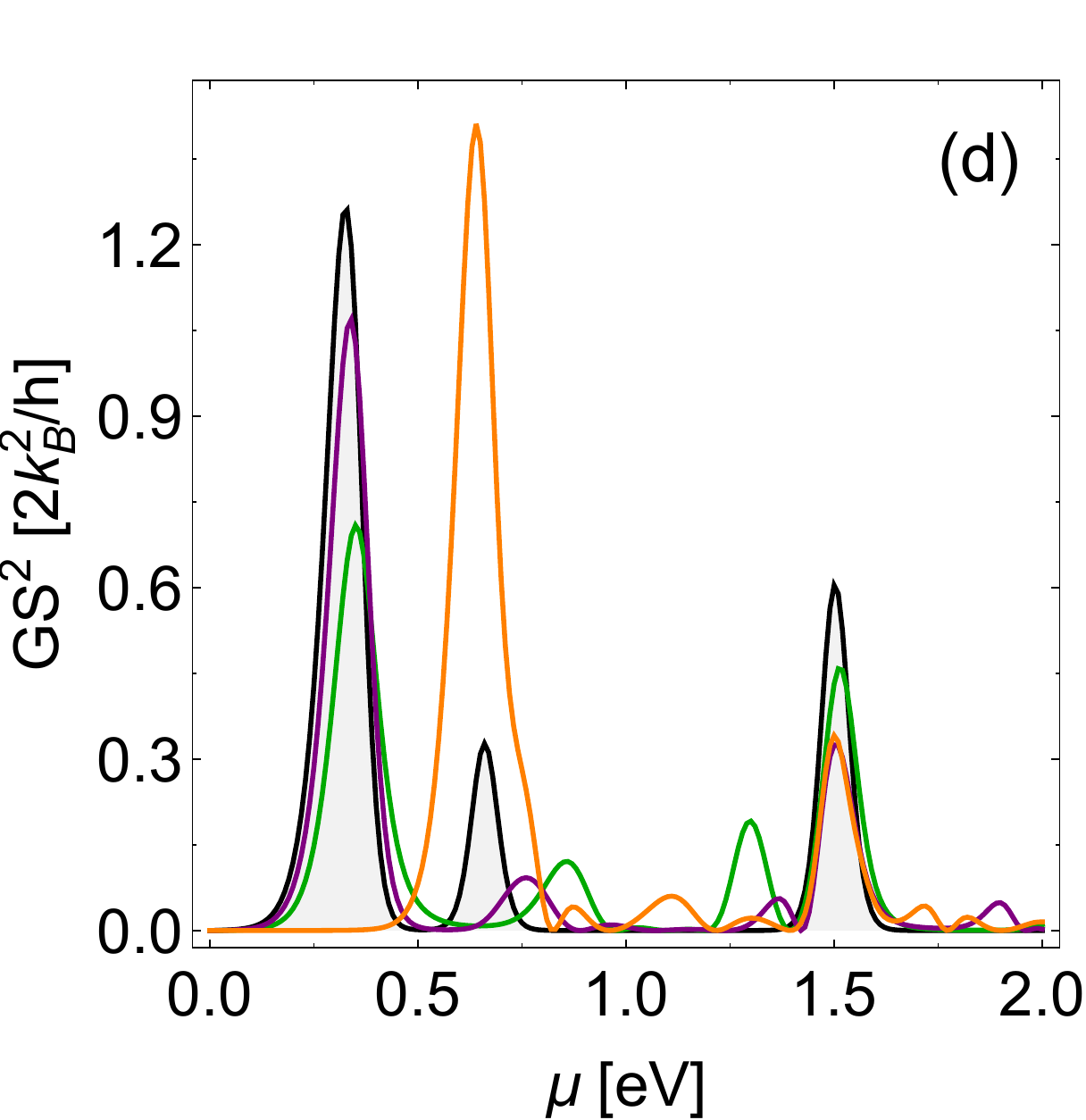}
\caption{Thermoelectric transport coefficients: (a) Seebeck coefficient $S$, (b) electrical conductance $G$, (c) electronic 
thermal conductance $\kappa_e$, and (d) power factor $P$, as functions of the chemical potential $\mu$ for a single $13$-AGNR held at room temperature. Black color corresponds to no defects and other color convention is the same as in 
Fig.\,\ref{fig:results_td_v}.}
\label{fig:results_vacancy}
\end{figure}

So far, we have considered the thermodefect voltage only. In order to understand these results, we next plot in Fig.\,\ref{fig:results_vacancy} the Seebeck coefficient $S$, electrical conductance $G$, electronic thermal conductance $\kappa_e$, and thermoelectric power factor $P$ as functions of the chemical potential $\mu$ in a single $13$-AGNR. We note that, unlike the thermodefect junction of two GNR components under a temperature gradient, we now consider 
the properties of a single GNR considering the constant chemical potential $\mu$ for both the reservoirs 
attached to the two opposite ends of the single ribbon. Comparing the thermoelectric properties of defective ribbons 
with the pristine one gives an idea about the behavior of a thermodefect junction made by these GNRs. The 
longitudinal line vacancies preserve the particle-hole symmetry, so the figures show only the results for electron 
doping, \ie positive chemical potential values. The zero of $\mu$ corresponds to the charge neutrality point, 
above (below) which the electrons (holes) are the dominant charge carriers. 

In Fig.\,\ref{fig:results_vacancy}(a), we see that the Seebeck coefficient for the pristine ribbon shows a dip for low 
values of the chemical potential $\mu$. But, notably, the values of $S$ in the presence of asymmetrical and edge longitudinal 
vacancies almost do not change from that of the pristine AGNR. However, the AGNR with the symmetrical longitudinal 
line vacancy has a considerably enhanced value of $S$ compared to the other two defect conditions. 
Thus, when a symmetrical longitudinal line vacancy is present, the Seebeck spectrum changes considerably. The differences between the Seebeck profiles for the different configurations is directly reflected in the thermodefect voltage shown in Fig.\,\ref{fig:results_td_v}, following the prediction of Eq.\,\eqref{Eq4}. This includes the voltage peak at $V_{\text{TD}}\sim0.5$ eV, where the difference between the Seebeck values for the pristine and symmetrical configurations also reaches its maximum.

Note that, we in Fig.\,\ref{fig:results_vacancy}(a) only plot for the regime: $0<\mu<1$ eV, as $S$ dies out at higher $\mu$ values for all defect configurations. 
This is due to the occupancy variance window (\ie derivative of the Fermi function  
$f(\varepsilon,\mu,T)$ with respect to the dimensionless chemical potential, $\mu/k_BT$) in the transport integral 
becomes narrower at higher $\mu$. This forces the transport integrals to decrease for any order of energy moments. However, the integrals with higher moments decrease faster than the lower ones. Therefore, $I_1/I_0$ decays rapidly leading to a strong decrease in the Seebeck coefficient at higher chemical potentials. The Seebeck spectrum of the pristine AGNR also display some smaller oscillations in the high chemical potential regime causing sign changes. These small oscillations can shift to other locations along the chemical potential axis with the change of the width of the AGNR\,\cite{geo4, tsegrap}. 
The sign changing behavior of the thermodefect voltage occurring in the higher chemical potential regime can be understood quantitatively from these oscillations of the Seebeck coefficients, more specifically, where the Seebeck 
coefficients of the pristine and defective AGNRs cross each other, see Fig.\,\ref{fig:results_vacancy}(a). Whenever 
the magnitude of the Seebeck coefficient of the pristine GNR is smaller than the defective one, a negative voltage is induced and vice versa, following the relationship between the Seebeck coefficients and the thermodefect voltage in Eq.\,(\ref{Eq4}). However, this oscillatory behavior of the voltage profiles has no practical importance because they tend to occur at high chemical potentials where the thermodefect voltages drop close to zero.

Moving on to the electrical conductance $G$ in Fig.\,\ref{fig:results_vacancy}(b), we see a regular stepwise pattern 
for the pristine AGNR consistent with earlier work\,\cite{therdefposit}. As soon as we put in a longitudinal vacancy, it 
starts deviating from the plateaus and resonances appear. The latter occur due to quantum interference effects caused 
by the vacancies in the AGNR\,\cite{PhysRevB.86.214206}. The behavior of the thermal conductance $\kappa_e$ of the ribbon is similar to the electrical conductance profiles, as shown in Fig.\,\ref{fig:results_vacancy}(c). The behavior of both the 
electrical and thermal conductances in the presence of the defects thus provides information on how the transport properties are affected by the defects. 

Finally, we compare the power factor $P$ between a pristine AGNR and those with different line vacancy defects in Fig.\,\ref{fig:results_vacancy}(d). The existence of large peaks for various chemical potentials indicates the range of 
$\mu$ for which our thermodefect junction can give the maximum power output\,\cite{introtebook}. The harvestable voltage range occurs to be at lower chemical potentials corresponding to large voltage values combined with the finite (non-vanishing) conductance. As a matter of fact, the power factor captures the chemical potential ranges where conductance and Seebeck coefficient are both large enough to get the maximum power. From the power factor analysis we can 
find the optimal conditions for $\mu$ that ensure the operation of a thermoelectric device with maximum output.

Till now, we have discussed only the longitudinal vacancies for the defective ribbon. We also investigate the thermodefect
voltages for our thermodefect junction when the defective ribbon has single, double, edge, antidot or transverse line vacancy, shown in Fig.\,\ref{fig:defects}(a)-(d) and (f), and present their maximum voltage values in Table\,\ref{table}, obtained for the same range of the chemical potential as used in Fig.\,\ref{fig:results_td_v}. For these vacancies, changing the position of the defects inside the GNR along both the transverse and longitudinal directions (with respect to the 
transport direction) does not affect the thermodefect voltages much, except the edge defects. This occurs due to the lattice symmetry along the transport direction is not changing between different vacancies. We note that the transverse line vacancy may be energetically unstable due to open tips and strong affinity of carbon atoms.  However, we still explore it for the sake of comparison with the other possibilities.

In order to investigate how the thermodefect voltages are sensitive to the edge configurations of the ribbon and also the 
existence of the band gap (finite for semiconducting and zero for metallic), we choose two different ribbon widths 
$W=11$ and $13$ for metallic and semiconducting AGNRs, respectively, and $W=12$ for ZGNR. In Table\,\ref{table} we see that for single, double, edge, transverse line and asymmetric longitudinal vacancies, metallic AGNR is the most suitable one in order to get larger voltages, whereas the semiconducting AGNR shows a considerably higher voltage for a symmetric longitudinal line vacancy. Other types of vacancies show more or less similar and notably smaller voltage values. Thus, on the whole AGNRs is more effective compared to ZGNRs. 

As soon as a defect is introduced in a pristine GNR, it induces some quasi-localized states and destroys some extended states in the DOS spectrum, which makes the electronic structure different than the pristine one, as already studied in the literature\,\cite{therdef2,PhysRevLett.96.036801,PhysRevB.77.085408}. We confirm this by calculating DOS and 
the transmission spectra but avoid providing the plots in order to focus on our main results of the thermodefect voltage. Overall, which defect type produces larger voltage depends on the number of localized states around the Fermi level as well as their degree of localization. The voltage values can be compared by analyzing the DOS and the transmission spectra for the defective ribbons. 

We also compare our results for the thermodefect junctions with vacancies to that for the junctions with impurities 
replacing the vacancies. Vacancies have more dramatic effects on the thermodefect voltage than impurities. The voltage decreases significantly when a vacancy is replaced by an impurity, no matter its physical placement. We therefore do not include showing the impurity results in Table\,\ref{table} as they have relatively very low thermodefect voltages. 

Throughout this work, we keep the widths of the GNR small to minimize the computational load and maximize the 
voltage, since the smaller the ribbon width, the larger the size effects and the higher the voltages. This is because the performances of nanoscale thermoelectric devices are enhanced, in general, as a result of the DOS modifications by the quantum confinement effects\,\cite{te4}. Thus, keeping the defect types same, decreasing the widths of both the junction components causes a steady increase in the voltage amplitudes.

\begin{table}[t]
\caption{Maximum thermodefect voltages for various defects in thermodefect junctions for $\Delta T=10$\,K. Three different widths 
of the GNRs are chosen. The cases with highest thermodefect voltages are emphasized by bold font.}
\begin{tabular}{lccc}
\hline \hline
   Defect type    & \multicolumn{3}{c}{$V_{TD}^{max}$ \text{[mV]}} \\
            & 11-AGNR     & 12-ZGNR     & 13-AGNR     \\ \hline
Single Vacancy                   & 1.53        & 1.43        & 0.28        \\
Double Vacancy                   & 1.62        & 1.26        & 0.25        \\
Edge Vacancy                       & 1.52        & 1.20        & 0.76        \\
Antidot Vacancy                   & 0.23        & 0.36        & 0.59        \\
Transverse line Vacancy                 & 1.56        & 0.42        & 0.95        \\
Symmetric Longitudinal  & -0.31        & 1.75        & 5.28        \\
Asymmetric Longitudinal  & 3.06        & -0.64       & 0.58        \\
Edge Longitudinal             & -0.36        & -0.60       & -0.32       \\
\textbf{Stone-Wales}  & 1.38   & 1.80 & \textbf{14.5}  \\
\textbf{555-777}  & -0.27  & 1.69  & \textbf{17.0}      \\
\hline \hline
\end{tabular}
\label{table}
\end{table}

\subsection{Dislocations}
Next we turn our attention to thermodefect junctions with dislocations in the form of Stone-Wales and 
$555$-$777$ defects. For these crystallographic dislocations the lengths of all nearby bonds are affected 
and that in turn affects the hopping integrals\,\cite{hop2,PhysRevB.81.195420}. We therefore adjust our Hamiltonian in Eq.~\eqref{Eq1} by making the hopping parameter $t$ site-dependent, choosing values near these defects from the density functional theory results existing in the literature\,\cite{hop1,hop3}. Although the hopping values near the vacancies can change, we find the thermodefect voltage to be robust against small changes in hopping values near the defects.

\begin{figure}[t]
\centering
\includegraphics[width=0.45\textwidth]{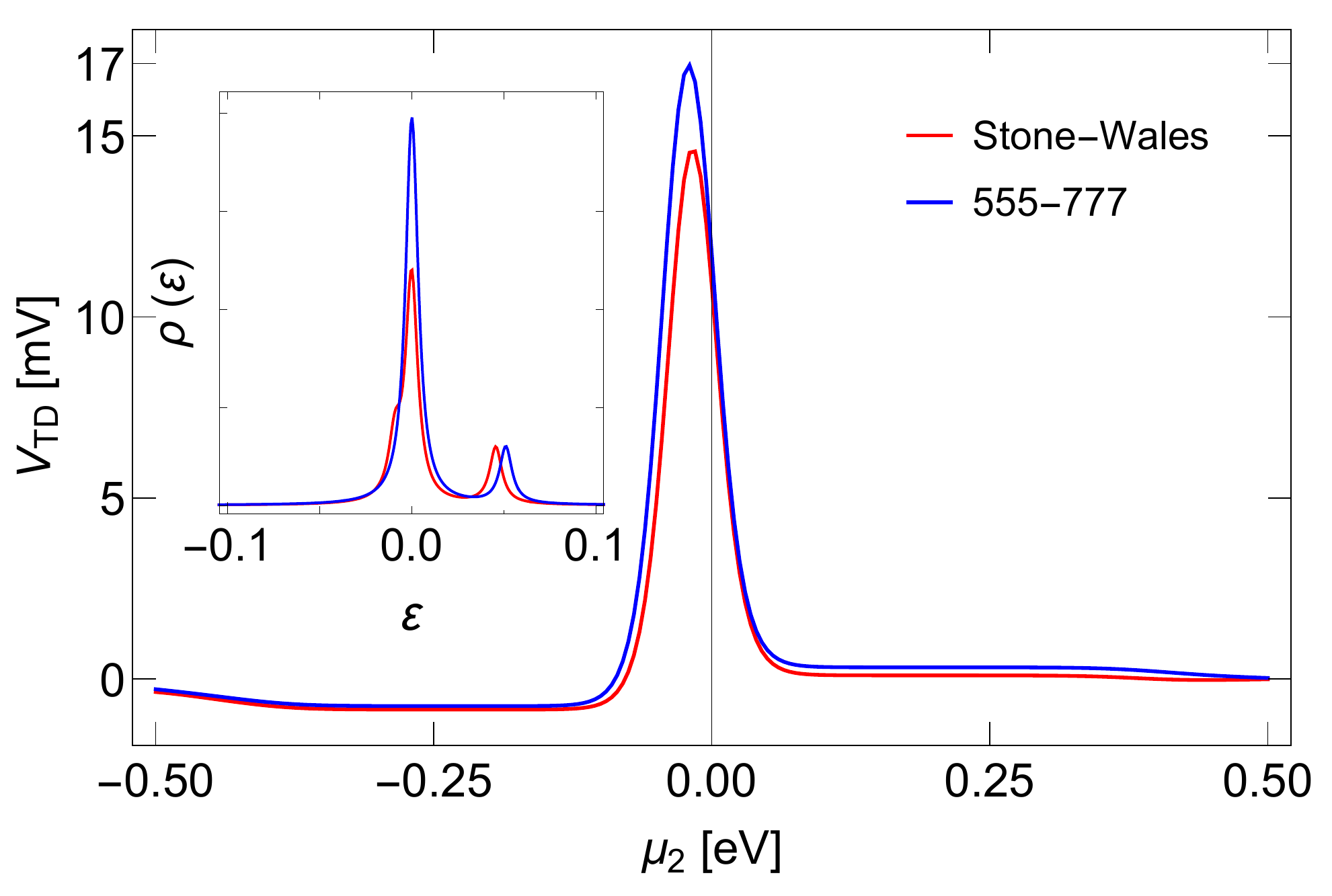}
\caption{Thermodefect voltage for $\Delta T=10$\,K as a function of the chemical potential at the common end 
$\mu_2$ for a thermodefect junction consisting of $13$-AGNRs. The red and blue colors represent the junctions 
with Stone-Wales and $555$-$777$ defects, respectively. Inset shows the DOS as a function of energy.}
\label{fig:results_td_defect}
\end{figure}

We show our results on thermodefect voltages for dislocations in Fig.\,\ref{fig:results_td_defect}  using a 13-AGNR (i.e.~semiconducting) junction. Here we plot the thermodefect voltage as a function of the chemical potential of 
the electrically connected end of the junction, within the range $-0.5\le\mu_2\le +0.5$ eV, as the thermodefect voltage dies out outside this range. We observe clear and large peaks at low $\mu_2$ for both dislocation defects, corresponding to the highest voltages highlighted (in bold) in Table\,\ref{table}. 

Interestingly, unlike vacancies, the voltage spectrum is asymmetric across the charge neutrality point ($\mu_2=0$) for GNRs with dislocation defects. We see a similar particle-hole asymmetry  exists in the DOS spectrum, plotted in the inset of Fig.\,\ref{fig:results_td_defect}. This asymmetry can be explained as follows.
In a regular honeycomb lattice, nearest neighbor hopping always occurs between 
different sublattices (\ie A to B or B to A, where A is denoted by blue and B is by orange color in Fig.\,\ref{fig:defects} respectively) and forms even-membered (even number of bonds) rings. Dislocations, like $555$-$777$ and Stone-Wales defects, however, break this sublattice symmetry around the defect regions\,\cite{PhysRevB.101.235116} and forms odd-membered rings (five or seven bonds) in a closed path encircling the dislocations\,\cite{PhysRevB.49.7697} by generating nearest neighbor hopping between the same sublattices (\ie from A to A or from B to B as shown in Fig.\,\ref{fig:defects} (g,h)).  It has previously been shown that propagation of the charge carrier's wave function through an odd-membered ring breaks the symmetry of the DOS with respect to the Fermi level and shifts the charge neutrality point\,\cite{PhysRevB.49.7697,PhysRevB.101.235116,PhysRevLett.96.036801,hopsame}. 
This happens because the hopping integrals between the same sublattices cause charge transfers between the localized and extended states, leading to a so-called self-doping process\,\cite{PhysRevB.73.125411}. As a consequence, the sublattice symmetry breaking surrounding dislocations leads to particle-hole symmetry breaking in the electronic spectrum clearly visible in the DOS and also, as we see in Fig.\,\ref{fig:results_td_defect}, for the thermodefect voltage.
On the contrary, for vacancies considered in the previous section, all  possible closed paths consist of even-membered sites. Therefore, particle-hole symmetry is preserved in both the DOS and the all transport properties. This also leads to the localized states induced by the vacancies being located exactly at the Dirac point to ensure charge neutrality\,\cite{PhysRevB.73.125411}.

The DOS plot in the inset of Fig.\,\ref{fig:results_td_defect} displays clear peaks for both dislocations. With the 13-AGNR being semiconducting, the DOS for the pristine case is zero for the same range of $\varepsilon$, making these DOS peaks even more prominent. We confirm that these DOS peaks represent quasi-localized states by calculating the transmission spectra. 
It is these quasi-localized states near the charge neutrality point that are also responsible for the large peak in the voltage spectrum in Fig.\,\ref{fig:results_td_defect}. Thus the DOS peaks and the associated quasi-localized states formed by dislocations clearly enhances the possibility of practical realization of the thermodefect effect.\,\cite{PhysRevLett.96.036801,PhysRevB.77.085408,local1}. 
In particular, we find that the quasi-localized states around zero energy in the presence of dislocations are much wider compared to the localized states formed by vacancies. This helps in obtaining enhancement in the thermodefect voltages for the dislocations. To explicitly verify this dependency, we split the transport integral into several energy regimes and find that the zero-energy states give the largest contribution to the observed voltage peaks. This result can also be inferred from the comparison of the Seebeck coefficients of the pristine and Stone-Wales or $555$-$777$ defective GNRs discussed later in this subsection. 

Next, we compare the maximum voltage values for these two dislocations for various widths of AGNRs and ZGNRs in Table\,\ref{table}. We see that $13$-AGNR stands out with considerably higher voltages for dislocation defects compared 
to the others. The reason for this can be attributed to the appearance of a wide resonance associated with quasi-localized states around the zero energy within the semiconducting gap of the 13-AGNR, whereas the presence of dislocations do not to the same degree affect the electronic structure with already existing flat bands in the ZGNR or the metallic state
in the AGNR. This is also in agreement with the difference in the voltage values depending on the types of defects \ie vacancies or dislocations, being affected by how many quasi-localized states are present around the Fermi energy. 

\begin{figure}[t]
\centering
\includegraphics[width=0.242\textwidth]{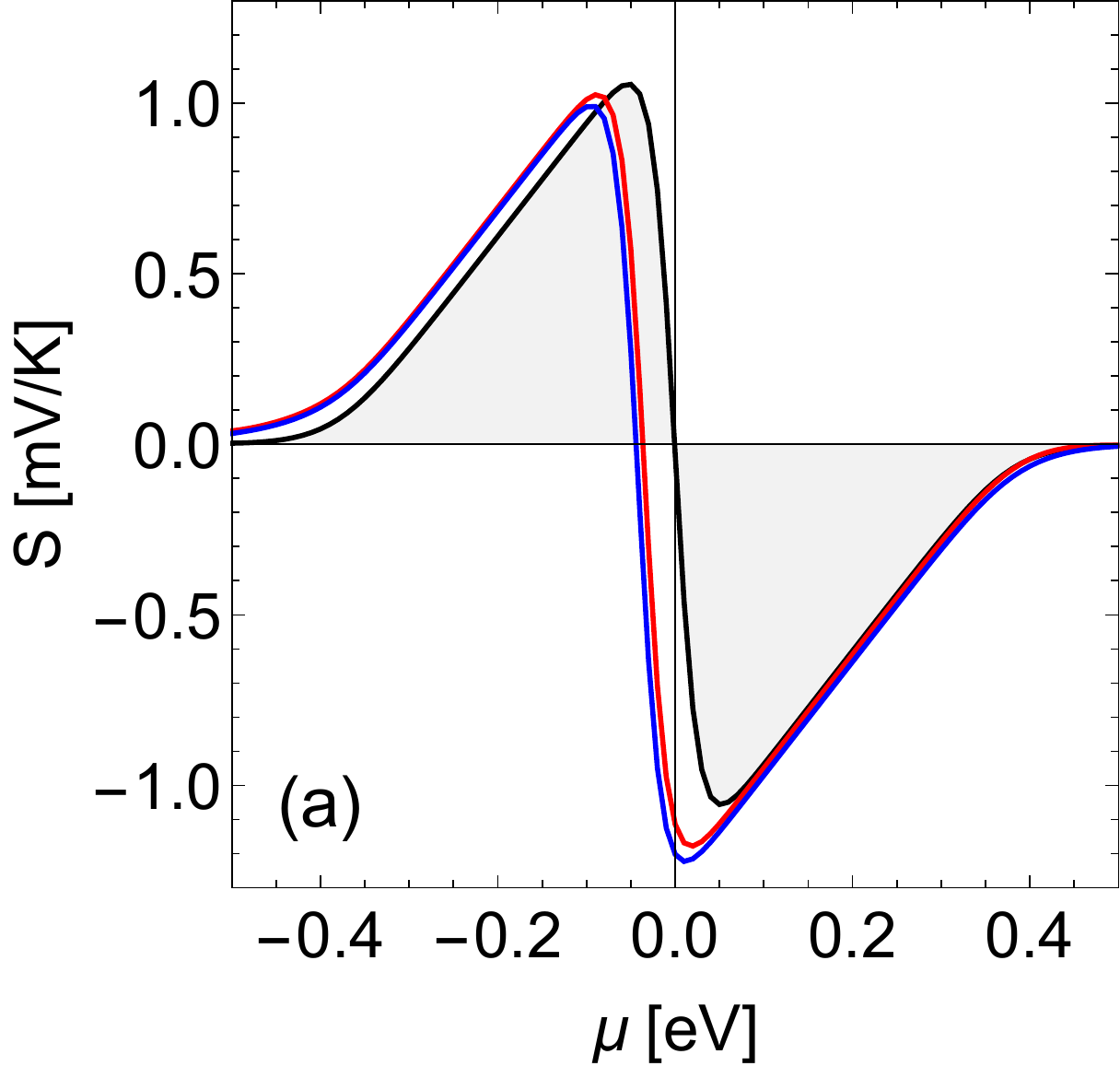}
\includegraphics[width=0.225\textwidth]{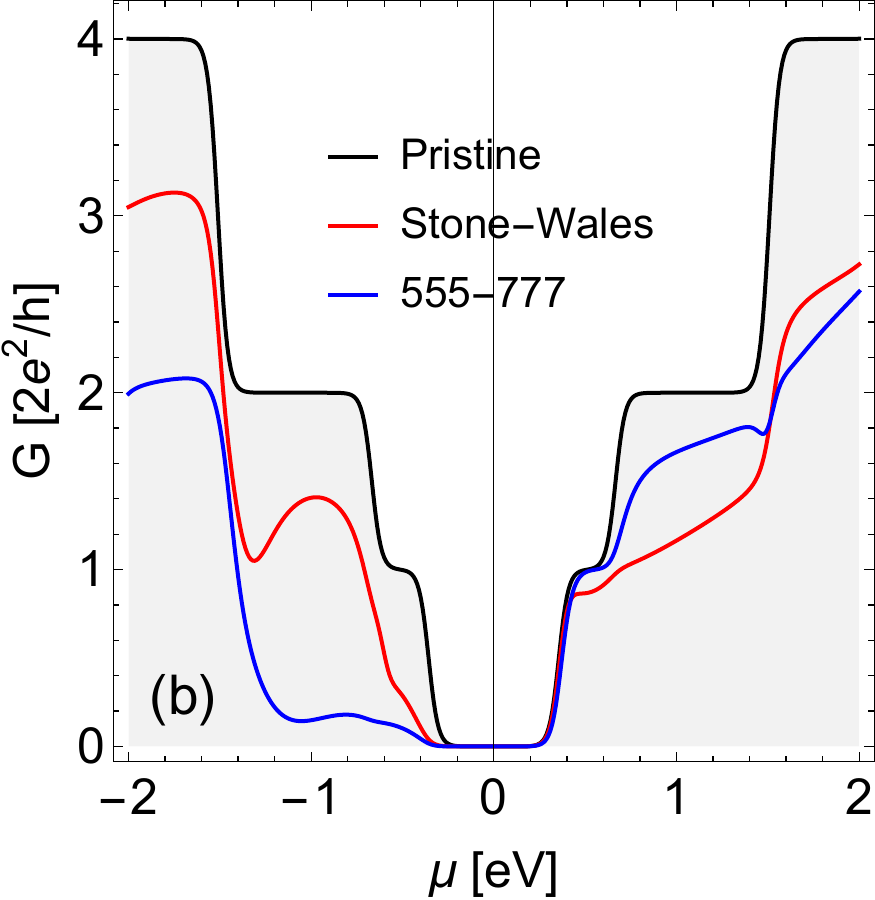}\,\,\,\,
\includegraphics[width=0.22\textwidth]{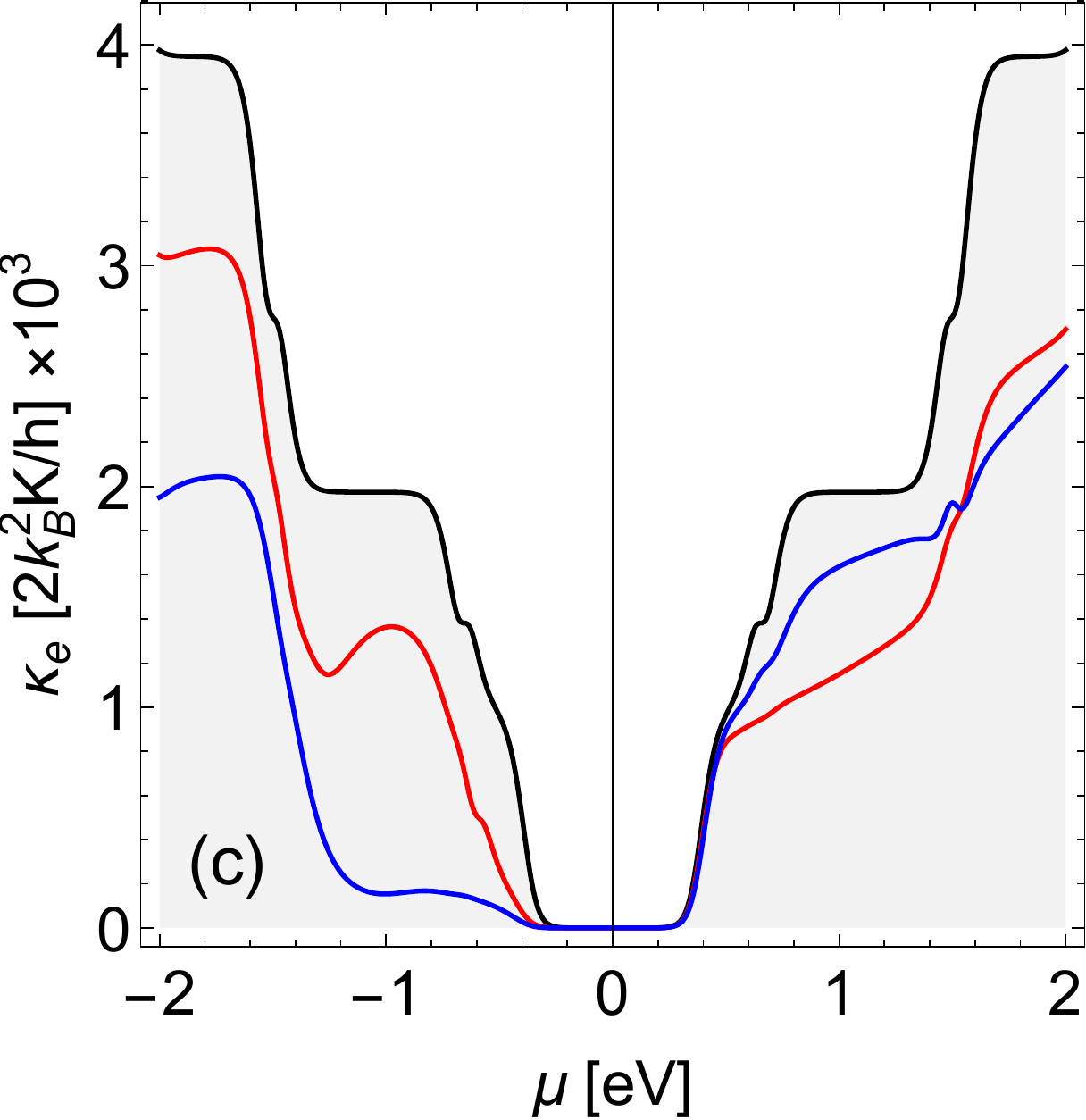}
\includegraphics[width=0.234\textwidth]{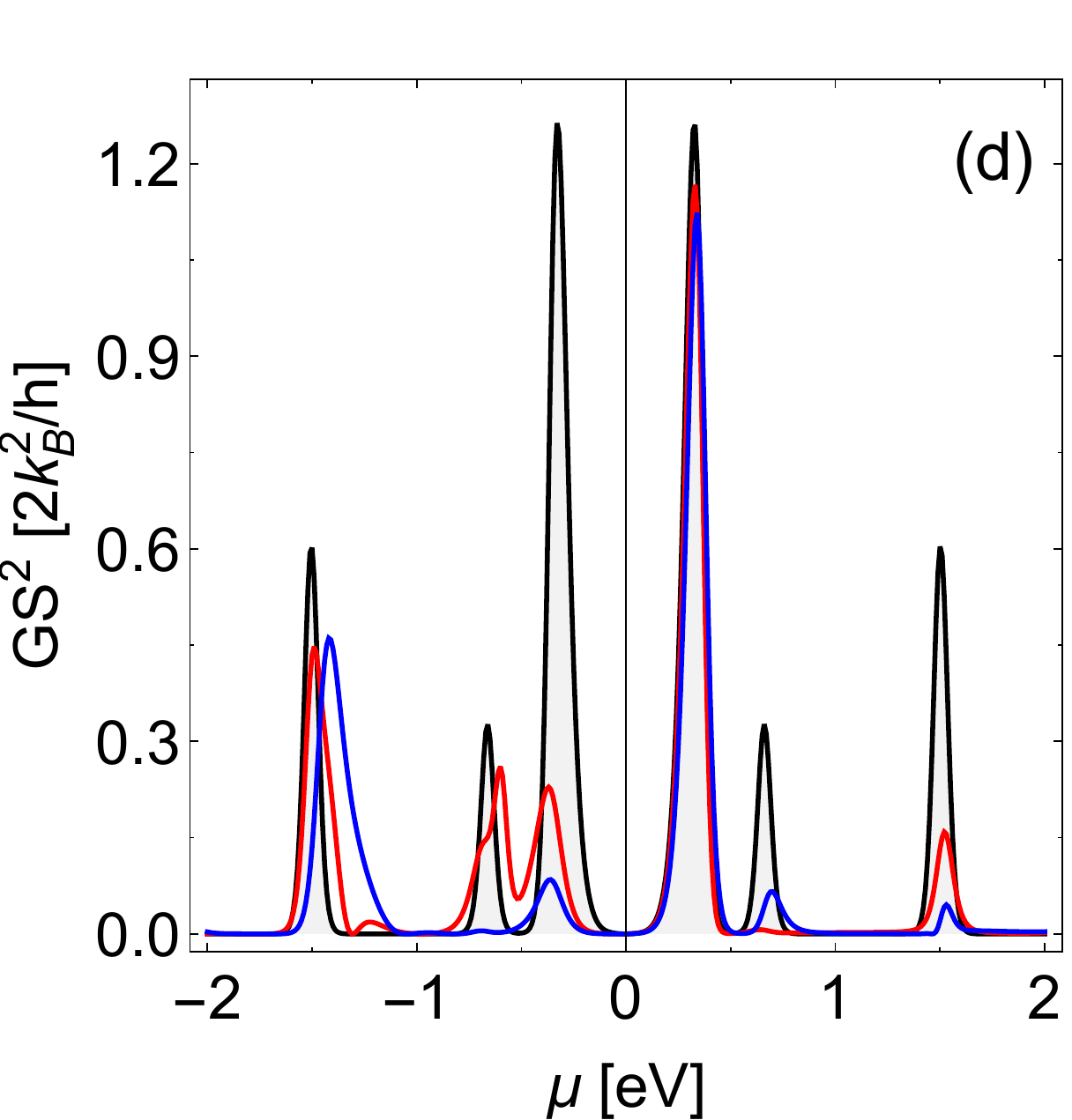}
\caption{Thermoelectric transport coefficients: (a) Seebeck coefficient, (b) electrical conductance, (c) electronic 
thermal conductance, and (d) power factor,  as functions of the chemical potential $\mu$ for a single $13$-AGNR at room temperature. The black, red, and blue colors represent no defect, with Stone-Wales and $555$-$777$ defects, respectively.} 
\label{fig:results_defect}
\end{figure}

To further understand the thermodefect voltages for dislocation defects we compare in Fig.\,\ref{fig:results_defect} the thermoelectric transport coefficients for a single pristine GNR (black curve) with that of single defective AGNRs with Stone-Wales or $555$-$777$ defects (red and blue curves, respectively). We observe that the transport properties of GNRs with Stone-Wales and $555$-$777$ defects are also asymmetric, just like the thermodefect voltage due to the broken particle-hole symmetry. Note that we in Fig.\,\ref{fig:results_defect} show the thermoelectric coefficients for the chemical potential within the range $\pm2$\,eV,  except the Seebeck coefficient, which is vanishingly small outside of $\pm0.5$\,eV. 
In terms of the Seebeck coefficient we observe sharp peaks around zero energy that are shifted in energy for GNRs with Stone-Wales and $555$-$777$ defects compared to the pristine case, see Fig.\,\ref{fig:results_defect}(a). This shift is due to the broken particle-hole symmetry and, combined with the pronounced low-energy peak structure, makes the difference in Seebeck coefficients between the defective and pristine GNRs particularly large around zero energy. Following the derivation in Eq.\,\eqref{Eq4} this directly explains the large thermodefect voltages around zero energy in Fig.~\ref{fig:results_td_defect}. 
The particle-hole asymmetry also appears in the electrical $G$ and thermal conductance $\kappa_e$ profiles, shown in Fig.\,\ref{fig:results_defect}(b) and (c), respectively. The deviations of the conductance profiles of the GNRs with the dislocations from that of pristine GNR are similar to that of the case of vacancy.  The asymmetry in the conductance spectra is transplanted also to the distinct peaks in the power factor $P$ in Fig.\,\ref{fig:results_defect}(d). 

To summarize, for dislocations we find that the larger the difference between the transport properties of the defective GNR and the pristine one, the higher the thermodefect voltages. In other words, the more asymmetry we have in the electronic spectrum in the presence of dislocation defects, the higher thermodefect voltages that can be obtained. This is a clear clue we can provide for the defect engineering to be used in the thermodefect junctions in reality.

\section{Conclusions} \label{IV}
We introduced the concept of defect-induced thermoelectric voltage, namely, thermodefect voltage, in a 
junction made of two identical GNRs but with defects present in one of the ribbons. We investigated this thermodefect
voltage for various defects: single vacancy, multi-vacancies (double, antidot, edge), as well as for dislocations and impurities.
For vacancies we found that the thermodefect voltage can be $\sim\,0.6$ mV/K for the case of longitudinal line
vacancies, whereas it can be as high as $\sim\,1.7$ mV/K when dislocation defects are used instead. For illustration we have shown results for mainly semiconducting AGNRs, as they are more effective in producing higher thermodefect voltages, but for completeness we also reported the thermodefect voltage values for AGNRs and ZGNRs with various widths.

In order to analyze the thermodefect voltages, we also studied the Seebeck coefficient, electrical conductance, electronic thermal conductance, and thermoelectric power factor of single AGNRs, both in pristine and defective conditions. Thermodefect voltages are related to signatures in the thermoelectric properties of the junction, the difference in the Seebeck coefficients between pristine and defective GNRs. 
Our results for the thermoelectric properties of a single GNR in the presence of defects are consistent with the sensitivity 
of the electric transport properties of GNRs to the configuration of defects as shown earlier in the literature\,\cite{therdef1,structdefect,Haskins2011,PhysRevB.91.035425,PhysRevLett.104.056801,roleofdef,reviewgnr}.
We also showed that vacancies and dislocations induce (quasi-)localized states by destroying the extended states of the GNRs. Combining these (quasi-)localized states with the particle-hole asymmetry produced by dislocations result in very large Seebeck coefficient differences between pristine and defective GNRs, which explain finding the largest thermodefect voltages in these setups.

Based on our results for the thermodefect voltage in many different GNR junctions we find that the performance of these thermodefect junctions is considerably better than traditional bulk systems and defects are thus able to make appreciable contributions for the enhancement of thermoelectric properties\,\cite{te3,tebook2014}. Depending on the defect type, the thermodefect voltage is even of the same order as in nanoscale bipolar thermoelectric junctions\,\cite{tebook2013}. 
It may in fact be possible to double the thermodefect voltage in a bipolar junction where both $p$ and $n$ components of 
the junction have exactly the same defects.

Finally, we emphasize that our concept of thermodefect voltage generation can be implemented in other 
nanostructured materials, beyond GNRs. Our thermodefect junction should be easy to implement experimentally. It additionally provides a varied and rich way 
of generating thermoelectric voltage owing to the existence of abundant defect types that can be engineered 
in the junction. Thus, the thermodefect effect proposed in this work may open up new efforts to further enhance 
the thermoelectric properties of nanostructured materials.

\acknowledgments{ABS and PD acknowledge the financial support from the Knut and Alice Wallenberg Foundation 
through the Wallenberg Academy Fellows program and the Swedish Research Council (Vetenskapsr\aa det Grant 
No. 2018-03488). AS wishes to extend his gratitude for the gracious hospitality and collaboration in Materials Theory Division, Department of Physics \& Astronomy, Uppsala University.}

\bibliography{bibfile}{}
\end{document}